# Cross Layer Implementation of Key Establishment and Configuration Protocols in WSN

Özgür Sağlam[1] and Mehmet E. Dalkılıç[2]

[1]Arçelik A.Ş. Electronics Plant R&D Department
Beylikdüzü, İstanbul, Turkey

[2]International Computer Institute, Ege University
Bornova, İzmir, Turkey

## Abstract

Security in Wireless Sensor Networks (WSN) can be achieved by establishing shared keys among the neighbor sensor nodes to create secure communication links. The protocol to be used for such a pairwise key establishment is a key factor determining the energy to be consumed by each sensor node during the secure network configuration. On the other hand, to achieve the optimum network configuration, nodes may not need to establish pairwise keys with all of their neighbors. Because, links to be established are defined by the network configuration protocol and as long as the network connectivity requirements are satisfied, number of links to be secured can be limited accordingly. In this sense, key establishment and network configuration performances are related to each other and this cross relation should be taken into consideration while implementing security for WSN. In this paper, we have investigated the cross layer relations and performance figures of the selected randomized pre-distribution and public key based key establishment protocols with the configuration protocol we proposed in a separate publication. Simulation results indicate that total network configuration energy cost can be reduced by reducing the number of links to be secured without affecting the global network connectivity performance. Results also show that the energy and resilience performances of the public key establishment can be better than the key pre-distribution for a given set of network configuration parameters.

## 1. Introduction[*]

Security in Wireless Sensor Networks (WSN) is important for the applications where the confidentiality and integrity of the sensed data to be analyzed are critical. Secure data transfer among wireless sensor nodes can be achieved by securing each of the links in the communication path via a secret key to be used for the message encryption. Establishing shared keys among sensor nodes after the deployment is a main research area in the literature.

Due to the limited power and computational resources of the WSN, literature has many proposed key establishment protocols that are mostly based on the key pre-distribution methods which do not require a lot of computation [1]. Among these methods, [2] is the first study proposing random key pre-distribution as an alternative to public key cryptography for pairwise key establishment. This work is known as Basic Scheme [1] and is the base for other probabilistic key establishment methods. The study [3] increased security with *q*-Composite scheme for small scale attacks but it suffers from large scale attacks since the network is completely compromised after a certain amount of nodes have been captured. [3] also proposed Multipath scheme and Random Pairwise scheme to increase resilience. However, Multipath scheme considerably increases the communication cost and Random Pairwise scheme reduces the scalability [1]. Another study [4] improves the communication cost by usage of polynomials but it increases the computation cost a little. [4] has a better resilience characteristic than [2] until a certain amount of nodes have been captured but when the threshold is reached the whole network is compromised [1]. Hierarchical method proposed in [5] uses polynomial based key pre-distribution proposed in [4] but it assumes exclusive nodes (Cluster Heads) to have a direct communication link with the sink and other cluster heads. This assumption makes the network to be vulnerable to single point of failure. On the other hand, if exclusive nodes are not assumed then this implementation is not practically possible for ordinary nodes since a wireless sensor node radio range is about 14 m when it is placed on the ground [6] which does not provide an effective cluster configuration with the direct links. Deployment knowledge based method [7] improves [2] but it increases the system complexity [1]. Besides, since there are assumptions with the deployment, [7] can not be applied to networks where there is no knowledge of the deployment.

On the other hand, eventhough the public key cryptography (PKC) methods RSA (Rivest, Shamir, Adleman) [8] and ECDH (Elliptic Curve Diffie-Hellman) [9] are accepted as they are expensive operations for a wireless sensor node, there are studies in the literature [10][11][12][13] evaluating the PKC based key establishment methods in WSN. In [10], RSA and ECDH energy consumption figures are investigated on Mica2dot wireless sensor nodes and [11] has extended this work to Mica2, MICAz and TelosB to investigate the effects of the RSA and ECC energy costs to sensor nodes lifetime. In [11] it has been concluded that the energy costs of asymmetric key establishment and even signature operations are not that critical, especially for Elliptic Curve Cryptography (ECC), compared to the total energy resources that these sensor nodes have. In [12] a configurable library, namely TinyECC, as an implementation of ECC for WSN has been proposed. TinyECC operates on a standard operating system TinyOS [14] which is suitable for common sensor platforms. A most recent study [13] has proposed a method which is an optimized implementation of ECDH key exchange for MICAz motes which also runs on TinyOS [14]. In [13] it has been figured out that the ECDH key exchange operation on the elliptic curve over a 192-bit prime field would have the key exchange energy cost of 57mJ per node which allows 117.000 key exchange operations for each node before running out of battery [13].

---

[*]An abbreviated version of this paper appeared in [15].

The main purpose of these key establishment protocols is to establish shared keys among sensor node pairs that might communicate and link to each other after the deployment to produce a connected network. From the network configuration point of view, some of these protocols [2][3][4] have been evaluated independently from the underlying network structure. However, if the network configuration requirements are also considered, assumptions made for the analysis of these protocols may not be applicable or their costs may not be acceptable for some WSN applications. For example, high node degree assumptions necessary for the key pre-distribution protocols [2][4] may not be practical due to the wireless medium efficiency. Besides, the high node degree requirement increases the total system cost per square area and this may not be acceptable for some WSN applications targeting large area coverage. Conversely, since the resources are limited, any limitation in the network configuration may affect the security level of the system which may require a configuration change in the key establishment protocol. Moreover, this change may not be enough to provide the desired security level.

On the other hand, cross layer relations of the key establishment and network configuration protocols may affect the system performance in a positive manner. For example, there can be some reduction in key establishment costs since the number of links to be established during the network configuration may be lower than the expected number of secured links assumed for the analysis of key establishment protocols. Therefore, building the cross layer relations with the network configuration protocol is necessary to evaluate the real performance of key establishment protocols.

In this paper we have investigated the cross layer relations and performance variations of the selected key establishment protocols together with the network configuration protocol proposed in [16] This network configuration protocol is a self organizing multi-hop clustering protocol which has been developed for homogenous WSNs to have maximum network connectivity with minimum node density (it provides the global connectivity of 98% with the minimum node degree of 7 [16]). This low node degree feature is important since it reduces the total system cost per unit area and eliminates the drawbacks of wireless medium limitations of dense networks stated in [17]. The protocol does not assume special sensor nodes having exclusive features like processing power, energy or extended radio range. Since it operates well with low node densities, to have the total network deployment cost per unit area as low as possible, the key establishment protocol to be implemented should provide the desired neighbor connectivity by increasing the probability of establishing secure links among neighbors as much as possible. Regarding these constraints, the key establishment protocols whose cross layer performances have been analyzed in this paper are based on key pre-distribution and PKC. The reference key pre-distribution protocol selected is the BS (Basic Scheme) [2] since it has been accepted as flexible, efficient, and fairly simple protocol while also offering good scalability [1]. We have selected Elliptic Curve Diffie-Hellman (ECDH) protocol for the PKC analysis since ECC is the most cost effective PKC solution suitable for WSN [11].

In this work, we have proposed three different link key setup procedures namely straight, reactive and proactive to establish the cross layer relation with the network configuration protocol by using its state control mechanism. Simulation results indicate that the cost of the key establishment protocols in WSN can be improved by controlling the number of links to be secured per sensor node, while keeping the global connectivity performance of the network configuration protocol at an acceptable level. Then we have analyzed the performance trade-offs of both key establishment protocols BS and ECDH implemented on the network configuration protocol [16]. The overall storage, computation, communication cost comparisons and resilience performances indicate that for the lowest physical node degree needed for 99% network connectivity, ECDH is a better choice if large scale networks are considered. For the small scale networks BS would be preferred but ECDH can still be a choice over BS since the total configuration energy costs are comparable. As the physical node degree increases then the energy cost of BS is becoming lower because of the reduced communication bandwidth usage. However ECDH can still be a preferred choice if the network resilience is the main concern. Finally we have analyzed the authenticated key establishment costs of signature schemes RSA [8] and ECDSA (Elliptic Curve Digital Signature Algorithm) [9] together with ECDH.

The rest of the paper is organized as follows. In Section 2, background information has been given. In Section 3, cross layer implementations of key agreement and network configuration protocols have been provided. In Section 4, simulation results have been given and conclusions are drawn in Section 5.

## 2. Background

In this section brief information on the protocols under concern has been given; the network configuration protocol proposed in [16], key establishment protocols BS [2] and ECDH [9].

2.1 Network Configuration Protocol

The network configuration protocol proposed in [16] is a multi-hop self organizing clustering protocol which organizes the network in a spanning tree form with an iterative heuristic. This configuration protocol is suitable for WSN requiring low node degrees starting from 7. Protocol basically starts with an initiator and grows with only local decisions based on the discovered neighborhood status. Once the first cluster is created, border members select new initiators for the configuration of new clusters. Among the responded nodes, one with the closest proximity and suitable neighborhood status is chosen as the new initiator. Then this node starts new cluster extension.

During the configuration every node in the network runs the same protocol state machine. There are five states for a sensor node which also define the node types namely, *floating, initiator, node, gateway_c* and *gateway*. At the beginning all the nodes are in *floating* state and they wait for an invitation from one of their already configured neighbors. A simplified state machine of this protocol is provided in Fig. 1.

State control mechanism of the network configuration protocol is as follows: In the *floating* state, a node may receive a configuration message from one of its already configured neighbors. After a configuration message is received any *floating* node first changes its state and starts neighbor discovery operation immediately. To discover its neighbors, first broadcasts a polling message and waits (one time interval) for the replies from its one hop neighbors, then it counts its neighbors and takes the corresponding action depending on the configuration message it received. For example, if this message is sent for starting a new cluster, then the state of the receiving node is changed to *initiator*. If the configuration message declares that the receiving node is an inner cluster member, then this node changes its state to *node* and continues extending the cluster with only its *floating* neighbors if they exist. If the receiving node is a cluster border member, then it changes its state to *gateway_c* and starts gateway decision algorithm defined in [16]. If this node finds itself as the best candidate among the other gateway candidates, then it changes its state to *gateway* and tries to select one of its *floating* neighbors as the initiator of the next cluster to be extended. Otherwise the state is changed to *node* and this unit stops its configuration process. This iteration stops when there is no further expansion is possible. When the protocol completes, an inter-cluster spanning tree is generated where the clusters created are the nodes of this spanning tree.

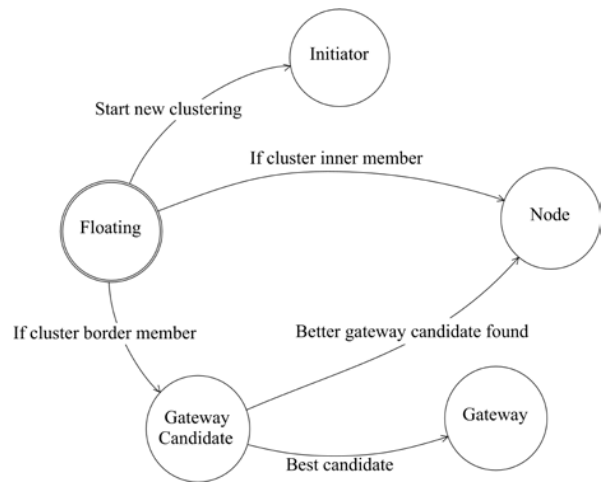

Fig. 1 Finite state machine of the network configuration protocol [16].

2.2 Basic Scheme (BS) [2]

In this protocol, each node randomly picks a subset of keys (i.e. key ring) from a large key pool and any pair of nodes can establish a secure connection if they share at least a common key in their key rings. The basic protocol operation consists of three phases. In *pre-distribution* phase a set of keys randomly selected from the key pool is stored in each sensor node, before the deployment. In *shared key discovery* phase, after the sensor nodes are deployed, each node checks its neighbors to see if they have a common key in their key rings. In *path key discovery* phase, if the originator does not have shared keys with some of its neighbors, it asks for path keys from the neighbor nodes that have a shared key with it. The path keys received are sent to those neighbors having no shared key with the originator and secure node connectivity is increased further.

The probability $p$ that two nodes share at least one key in their key rings of size $k$ chosen from a given pool of $P$ keys is defined in [17] as:

$$p = 1 - \frac{(1-k/P)^{2(P-k+1/2)}}{(1-2k/P)^{(P-2k+1/2)}} \quad (1)$$

If we define $d$ as the physical node degree of the network, then the expected number of neighbors establishing link key in shared key discovery is $pd$ and it is $(1-p)(1-(1-p^2)^d)d$ for path key discovery. Here we assume that only single hop neighbors are used for path key discovery.

Based on this probabilistic scheme, the local connectivity and global connectivity relations first investigated in [17]. This work states that the local connectivity requirements of random graphs can be reduced if a global connectivity of less than 100%, for instance 98% is targeted. In [17], there are also two methods defined for path key discovery phase of the randomized key pre-distribution protocols namely *cascade-off* and *cascade-on counting*. With the *cascade-on counting*, all the nodes establishing a path key with the originator in one round of the path key discovery are counted as the new shared key neighbors of this originator in the following round. This operation terminates when there is no non-secured neighbor left or no new neighbor having established a path key with the originator in the last round of the path key discovery. With the *cascade-off counting*, there is only one round of path key discovery and the neighbors who have established a session key with the originator in the shared key discovery phase are used for the path key setup.

2.3 Elliptic Curve Diffie-Hellman (ECDH)

ECDH public key establishment protocol requires two point multiplications and one message per unit for the shared key establishment. The first point multiplication is for calculating the public key $Pu$ from the unique private key $P$. After having $(P, Pu)$ pairs, each node $u$ and $v$ starts messaging for session key establishment, $u$ sends its public key to $v$, $u \rightarrow v: \{Pu_u\}$ and $v$ sends its public key to $u$, $v \rightarrow u: \{Pu_v\}$. Then, each node computes the shared key $K_{link}$ via the second point multiplication as:

$$K_{link} = P_u Pu_v, \quad K_{link} = P_v Pu_u \quad (2)$$

In addition, there are signature protocols RSA and ECDSA defined for authenticating the public keys to defeat the *man in the middle* attacks. Recent works indicates that implementation of these signature protocols on wireless sensor nodes are possible [10][11][12] and there are software libraries developed [12]. More generally, signature generation and verification operations of these protocols have different asymmetrical computation costs for each [11]. Generation of signatures is handled by a trusted party which can be called as Certificate Authority (CA). For this operation, public key of each node is signed by the private key of CA and the signed copy is loaded to the corresponding node. Then, each communicating node verifies the signature by using the public key of the CA to authenticate the public key of the sender. In WSN this CA can be the base station or any other powerful device in which all the signed copies of the public keys can be calculated and copied to the corresponding nodes before the deployment. After the deployment, the sensor nodes can verify any of its neighbor signatures by using the public key of the signing unit.

## 3. Cross Layer Implementations of Key Agreement Protocols with the Network Configuration

In this section we provide the design details of the cross layer implementations of key establishment protocols over the network configuration protocol proposed in [16]. We first detailed the implementation of the BS and then, the details of ECDH implementation have been given.

3.1 Implementation of Basic Scheme (BS)

Cross layer implementation of BS over the configuration protocol [16] requires additional communication messages in the neighbor discovery phase. The configuration protocol states that, any *floating* sensor node receiving a configuration message starts neighbor discovery operation. During this phase, states of the sensor nodes in the neighborhood are registered and cluster extension is managed with only the *floating* neighbors. Pre-distributed key rings are checked in this phase and the neighbors, sharing at least a key, participate in the next steps of the network configuration protocol.

Key pool size $P$ and key ring size $k$ of BS define the probability $p$ as in Eq. (1). Then, by changing these parameters, the number of neighbors that can be connected securely can be controlled. However, this affects the cost of storage and transmission.

Transactions for the basic BS implementation in neighbor discovery phase of the configuration protocol have been illustrated in Fig. 2. Here any configured sensor node $A$ starts neighbor discovery by broadcasting a polling message first (Fig. 2-I). This message includes $k_A$ which is the list of key indexes included in the key ring of *A*. Receiving neighbors $n_i$ (*i*=1,2,...,*d* ; *d* is the expected node degree) compare $k_A$ with their key rings $k_i$. If there is at least one key index match, then the corresponding neighbor replies to *A* with this number $idx_i$ (Fig. 2-II). This reply is the confirmation of the link key between *A* and the node *i*. The expected number of neighbors finding a common key after the shared key discovery is *pd*. So there are (1-*p*)*d* neighbors cannot find a shared key with $k_A$. These neighbors reply with the list of their key indexes $k_i$ (*i*=1,2,...,(1-*p*)*d*) which are to be used in the path key discovery phase (Fig. 2-II). After receiving all replies, *A* is able to register its neighbors according to their key sharing status. Depending on the probability *p*, if there are neighbors that could not established a link key, *A* starts path key discovery process through its *pd* secure neighbors which have a common key with *A*. In this phase, *A* broadcasts the list of key indexes $k_i$ (*i*=1,2,...,(1-*p*)*d*) received in shared key discovery phase together with the corresponding node id numbers $id_i$ (Fig. 2-III). After receiving this message, each of the *pd* secure neighbors check their key index lists and prepares a path key for each one if there is at least a common key found. Then, each path key is sent to *A* in two copies which has been referenced as $K_{PK}$ in Fig. 2. One copy is encrypted with the link key of *A* and the other is encrypted with the key shared with the corresponding key ring i.e. indirect neighbor (Fig. 2-IV). Each $K_{PK}$ pair has been associated with also the target neighbor *id* and its key ring index number for the link key establishment. In these reply messages, there can be multiple path keys $K_{PK}$ defined for a single neighbor $n_i$. In this case, *A* selects only one of them. Then, the corresponding path key $K_i$ is sent to the corresponding neighbor with its key index number $idx_i$ where $i \leq$ (1-*p*)*d* (Fig. 2-V). If *cascade-off counting* is selected, key setup process terminates at this point. Otherwise, if there are neighbors that still cannot establish a path key, then *A* starts a new path key discovery iteration starting from the step Fig. 2-III. New iteration includes also the last added neighbors in the previous path key discovery as the shared key members. This loop terminates when there is no unsecured neighbor left or no new neighbor added after the last path key discovery process (*cascade-on counting*).

Per node energy cost of this key establishment scheme can be divided into two parts as communication and computation costs. For the analysis, only the single hop neighbors for path key discovery is considered and *cascade-off counting* has been applied.

### 3.1.1 Communication Cost of BS

**Lemma 1:** In shared key discovery phase any sensor node $v_i$ (*i* = 1,2,3,...,*n*) makes a single broadcast and $E_{neighbor}$ unicasts, where $E_{neighbor}$ is the expected number of physical neighbors that are a single hop away from $v_i$.

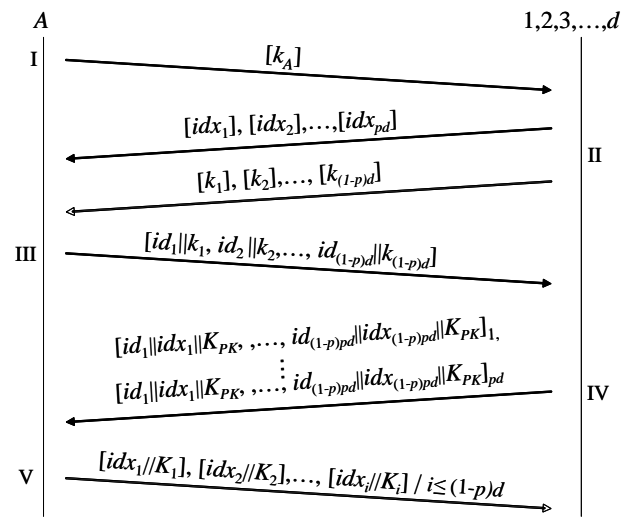

Fig. 2 Message transactions for BS.

**Proof:** At the beginning, each sensor node is in the *floating* state and waits for the configuration message from one of its already configured neighbors. In the network configuration protocol, a node changes its state from *floating* and starts neighbor discovery operation only after a configuration message is received. Consequently, any node $v_i$ receives a single valid configuration message and makes a single broadcast for neighbor discovery during the network configuration. As a result, any node $v_i$ receives a single broadcast message from each of its single hop neighbors ( $u_1, u_2, u_3, ..., u_{E_{neighbor}}$ ) and unicasts the replies including its state information. Each reply includes also the key ring indexes of the node if there is no shared key found in the configuration message. Total number of these reply messages that $v_i$ transmits varies with the expected number of physical neighbors which is $E_{neighbor}$. As a result, total number of transmitted messages from $v_i$ for neighbor discovery equals to 1+ $E_{neighbor}$ □

**Lemma 2:** In shared key discovery phase any sensor node $v_i$ (*i* = 1,2,3,...,*n*) makes a single broadcast and $E_{neighbor}$ unicasts, where $E_{neighbor}$ is the expected number of physical neighbors that are a single hop away from $v_i$.

**Proof:** At the beginning, each sensor node is in the *floating* state and waits for the configuration message from one of its already configured neighbors. In the network configuration protocol, a node changes its state from *floating* and starts neighbor discovery operation only after a configuration message is received. Consequently, any node $v_i$ receives a single valid configuration message and makes a single broadcast for neighbor discovery during the network configuration. As a result, any node $v_i$ receives a single broadcast message from each of its single hop neighbors ($u_1, u_2, u_3, ..., u_{E_{neighbor}}$) and unicasts the replies including its state information. Each reply includes also the key ring indexes of the node if there is no shared key found in the configuration message. Total number of these reply messages that $v_i$ transmits varies with the expected number of physical neighbors which is $E_{neighbor}$. As a result, total number of transmitted messages from $v_i$ for neighbor discovery equals to $1 + E_{neighbor}$ □

**Lemma 3:** In shared key discovery phase any sensor node $v_i$ ($i = 1,2,3,…,n$) receives $E_{neighbor}$ broadcast and $E_{neighbor}$ unicast messages.

**Proof:** From Lemma 1, each node $v_i$ makes a single broadcast for neighbor discovery. This message is received by each of $E_{neighbor}$ nodes ($u_1, u_2, u_3, ..., u_{E_{neighbor}}$) and these nodes unicast a reply message in return. In this case, there are $E_{neighbor}$ unicasts received by $v_i$. On the other hand, $v_i$ receives a single broadcast from each of its single hop neighbors $u_1, u_2, u_3, ..., u_{E_{neighbor}}$ which has been sent once for neighbor discovery. As a result, there are total of $E_{neighbor}$ broadcast messages received by $v_i$. □

**Lemma 4:** The total cost for messages transmitted from each node $v_i$ ($i = 1,2,3,…,n$) is calculated as below:

$$C_{M_{tx}} = C_{tx}(L_{poll\_sk} + L_{poll\_pk} + E_{neighbor}(pL_{poll\_reply\_s} + (1-p)(L_{poll\_reply} + p^2 E_{neighbor} L_{ask\_key} + L_{ask\_key}/2))) \quad (3)$$

($C_{tx}$: radio *transmission* cost per bit, $L_{poll\_sk}$: broadcast message length in shared key discovery phase, $L_{poll\_pk}$: broadcast message length in path key discovery phase, $L_{poll\_reply\_s}$: Length of a reply message for shared key information, $L_{poll\_reply}$: Length of a reply message for no shared key information and $L_{ask\_key}$: Length of unicast messages transmitted for path key establishment)

**Proof:** As it is given in Lemma 1, each node $v_i$ transmits a single broadcast message in shared key discovery phase and there is another broadcast message transmitted in path key discovery which includes the key rings of unshared key members. The bit length of these messages are $L_{poll\_sk}$ and $L_{poll\_pk}$ respectively. Each node $v_i$ replies to these broadcast messages received from $E_{neighbor}$ nodes after checking the shared key status with the received key ring indexes. Hence, $v_i$ sends only the shared key index to $pE_{neighbor}$ nodes and its key ring indexes of length $k$ to $(1-p)E_{neighbor}$ nodes. The bit length of these messages are $L_{poll\_reply\_s}$ and $L_{poll\_reply}$ respectively. Additionally, each $v_i$ unicasts its reply to maximum of $pE_{neighbor}$ broadcast messages with maximum of $p(1-p)E_{neighbor}$ path keys included in each reply. Here, each path key message length is $L_{ask\_key}$. Lastly, the length of the message for sending the encrypted path keys to $(1-p)E_{neighbor}$ neighbors is $L_{ask\_key}/2$. As a result, total unicast message length transmitted is:

$$E_{neighbor}(pL_{poll\_reply\_s} + (1-p)(L_{poll\_reply} + p^2 E_{neighbor} L_{ask\_key} + L_{ask\_key}/2)) \quad (4)$$

Since the total broadcasted message length is calculated as $L_{poll\_sk} + L_{poll\_pk}$, by using Eq. (4), total message transmission cost for a node is calculated as below:

$$C_{M_{tx}} = C_{tx}(L_{poll\_sk} + L_{poll\_pk} + E_{neighbor}(pL_{poll\_reply\_s} + (1-p)(L_{poll\_reply} + p^2 E_{neighbor} L_{ask\_key} + L_{ask\_key}/2))) \quad □$$

**Lemma 5:** The total cost for messages received by each node $v_i$ ($i = 1,2,3,…,n$) is calculated as below where, $p$ is the probability that two nodes share at least one key in their key rings and $n$ is the total number of nodes in the network. ($C_{rx}$: radio reception cost per bit)

$$C_{M_{rx}} = C_{rx} E_{neighbor}((L_{poll\_sk} + L_{poll\_pk}) + pL_{poll\_reply\_s} + (1-p)(L_{poll\_reply} + p^2 E_{neighbor} L_{ask\_key} + L_{ask\_key}/2)) \quad (5)$$

**Proof**: At first, each node $v_i$ in the network receives the broadcast messages $L_{poll\_sk} + L_{poll\_pk}$ defined in Lemma 3. In addition, $v_i$ receives reply messages from its neighbors in return to its previously transmitted broadcast messages. Firstly, it receives shared key list index ($L_{poll\_reply\_s}$) from $pE_{neighbor}$ neighbors and key ring indexes of length $k$ ($L_{poll\_reply}$) from $(1-p)E_{neighbor}$ neighbors after its broadcast for shared key discovery. In path key discovery phase, $v_i$ receives path keys ($L_{ask\_key}$) for at most $p(1-p)E_{neighbor}$ neighbors from each of $pE_{neighbor}$ neighbors. Finally, the path key ($L_{ask\_key}/2$) is received from each of at most $(1-p)E_{neighbor}$ neighbors. As a result, total unicast messages received is:

$$E_{neighbor}(pL_{poll\_reply\_s} + (1-p)(L_{poll\_reply} + p^2 E_{neighbor} L_{ask\_key} + L_{ask\_key}/2)) \quad (6)$$

Since the received broadcast message length is $E_{neighbor}(L_{poll\_sk} + L_{poll\_pk})$, by using Eq. (6), total message reception cost is calculated as below where $C_{rx}$ is the radio reception cost per bit:

$$C_{M_{rx}} = C_{rx} E_{neighbor}((L_{poll\_sk} + L_{poll\_pk}) + pL_{poll\_reply\_s} + \\ (1-p)(L_{poll\_reply} + p^2 E_{neighbor} L_{ask\_key} + L_{ask\_key}/2)$$

Then, by using the results of Eq. (3) and Eq. (5), the total messaging cost of each node $v_i$ ($i = 1,2,3,…,n$) for secure link establishment is calculated as below:

$$C_{M_{neighbor}} = C_{M_{tx}} + C_{M_{rx}} \qquad (7)$$

### 3.1.2 Computational Cost of BS

The computational costs of the BS are the encryption and decryption costs of path keys during path key discovery. During the path key establishment each node $v_i$ decrypts at most $(1-p)d$ path keys received from its shared key neighbors. In addition, each node performs encryption for each of $(1-p)d$ key rings received from each of $d$ neighbors with the probability of $p^2$ for sharing key with the key ring and the originator node. This encryption is made twice, one for the originator and the other for the node who needs the path key. Then, the total number of encryptions is calculated as $2p^2(1-p)d^2$ and the resulting cost is given as:

$$C_{computation} = (2p^2(1-p)d^2)_{enc} + ((1-p)d)_{dec} \qquad (8)$$

### 3.1.3 Total Energy Cost of BS

The message payload structure used in the energy cost analysis of the BS is:

$$A \to B : ID_A, ID_B, M, MAC\{ID_A, ID_B, M\}$$

In this structure, $ID$ is the network wide unique identification number of each sensor node. We have taken $ID$ length as 16 bits to handle the large scale networks (ie. > 512). $M$ includes the keys, key lists and corresponding $ID$ numbers attached and the length of this part is variable. Here, the MAC (Message Authentication Code) is calculated by using SHA-1 and it occupies 160 bits in the payload. The encryption/decryption protocol referenced is AES 128 which has the corresponding key length of 128 bits. For this analysis the key pool size is taken as $P = 10.000$ which corresponds to 16 bits key ring indexes. The energy cost equation is derived as a function of the key ring size $k$.

Before starting the analysis, we need to define the length of each messages used in communication cost analysis of BS. The $L_{poll\_sk}$ sent for the shared key discovery includes the indexes of the keys stored in $A$ and it has the length of $2k$ bytes. The $L_{poll\_reply}$ message is a reply to $L_{poll\_sk}$ and it includes the key ring index list of the sender if there is no shared key found. This message also has the length of $2k$ bytes. If there is a shared key, then the 16 bits index number of the shared key ($L_{poll\_reply\_s}$) is replied. $L_{poll\_pk}$ message sent in path key discovery includes the $ID$ numbers (16 bits) and key index lists of the neighbor nodes ($L_{poll\_reply} = L_{poll\_sk}$) sharing no key with $A$ after the shared key discovery. The number of these nodes depends on the probability $p$ and node degree $d$ as below:

$$M = L_{poll\_pk} = d(1-p)(L_{poll\_sk} + 2) \qquad (9)$$

Any neighbor node generating the path key, encrypts it with the key shared with the originator and the key shared with the target node. These two encrypted keys are then sent to the originator node. This message is $L_{ask\_key}$ and it includes two pieces of 128 bits of encrypted keys, 16 bits $ID$ of the target node and 16 bits index number of the key to be used for decrypting the encrypted path key in that node. The originator sends the corresponding encrypted copy (128 bits) to the target node for path key establishment together with the index number (16 bits) of the key to be used for decrypting the key in that node. Lengths of these messages are summarized in Table 1.

Table 1: Basic Scheme (BS) message lengths

| Symbol | Message Length (bit) |
|---|---|
| $L_{poll\_sk}$ | 16$k$ |
| $L_{poll\_reply}$ | 16$k$ |
| $L_{poll\_reply\_s}$ | 16 |
| $L_{poll\_pk}$ | $d(1-p)$ ($L_{poll\_sk}$ + 16) |
| $L_{ask\_key}$ | 2(16+128) |

Now we can calculate the total energy cost of the BS for a single node. In addition to the message types $M$ defined, each payload includes also $ID$ numbers of the sender/receiver and the MAC part which is 160 bits SHA-1 output of the $ID$s and $M$. Thus, the Eq. (3) and Eq. (5) can be rewritten as below:

$$C_{M_{tx}} = C_{tx}(16k + 192 + d(1-p)(16k+16) + 192 + \\ d(p208 + (1-p)(16k + 192 + p^2 d 480 + 336))) \qquad (10)$$

$$C_{M_{rx}} = C_{rx} d(16k + 192 + d(1-p)(16k+16) + 192 + \\ p208 + (1-p)(16k + 192 + p^2 d 480 + 336)) \qquad (11)$$

In [10] and [11], the Rx/Tx radio communication and AES 128 calculation costs of Mica2dot sensor nodes have been given and the reference values taken for the analysis are provided in Table 2 and Table 3 respectively. Here, the per bit energy costs given in Table 2 define the $C_{tx}$ and $C_{rx}$ costs included in our analysis.

Table 2: 868 Mhz Rx/Tx costs for Mica2dot

| Radio | Energy |
|---|---|
| $C_{rx}$ | 0.750 μJ/bit |
| $C_{tx}$ (5 dBm) | 1.984 μJ/bit |

Table 3: Symmetric computation costs for Mica2dot

| Computation Type | Energy |
|---|---|
| AES128$_{enc}$ | 1,62 μJ/byte |
| AES128$_{dec}$ | 2,49 μJ/byte |
| SHA-1 | 5,9 μJ/byte |

The total amount of AES 128 computations is defined in Eq. (8). Each of these computations is done for 128 bits = 16 bytes keys, so the total computation cost can be rewritten as below:

$$C_{computation} = (2p^2(1-p)d^2)16_{enc} + (2(1-p)d)16_{dec} \quad (13)$$

Then, by using the equations Eq. (10), Eq. (11), Eq. (12) and Eq. (13), the overall per node energy cost of BS for $v_i$ ($i = 1,2,3,…,n$) is calculated as below:

$$C_{configuration} = C_{M_{tx}} + C_{M_{rx}} + C_{M_{SHA-1}} + C_{computation} \quad (14)$$

### 3.1.4 Cross Layer Implementation of BS

In BS, securing all the physical neighbors for each sensor node would give the highest network configuration performance. However, the communication and computation costs are higher in this case. Instead, only the required neighbors which are in *floating* state can be selected for establishing secure links while extending clusters. This reduction would improve the communication costs and hence the computation costs because of the reduced MAC computations. We have defined three different methods for selection of links to be secured as *proactive*, *reactive* and *straight*.

--With the *proactive* selection, the links to all the physical neighbor nodes are tried to be secured in shared key and path key discovery phases (uses *cascade-on counting*).

--With the *reactive* selection, the links with only *floating* neighbors are tried to be secured in shared key discovery. If all of them cannot be secured, then the path key discovery process is applied for the remaining *floating* neighbors just once (uses *cascade-off counting*).

--*Straight* selection only considers the neighbors having shared keys discovered in the shared key discovery phase and path key discovery does not run in this mode.

The algorithm of this cross layer implementation of BS is given in Table 4. This algorithm is included in the neighbor discovery part of the configuration protocol. Since the algorithm defines the links to be securely established during the neighbor discovery, after its completion, the network configuration protocol continues its operation with the neighbors having shared keys. Simulation results of these implementations and their impacts on security and network configuration performance have been provided in Section 4.

Table 4: BS cross layer implementation algorithm

**Assumptions:** Network is $G = (V;E)$, there is no node failure during configuration
1: BEGIN:
2: $i \leftarrow 0$; $j \leftarrow 0$; *new_neighbor_added* $\leftarrow 0$
3: BROADCAST *poll* with $k_v$ for neighbor discovery and start *neighbor discovery timer*
4: RECEIVE *poll_reply* from $u$
5: **if** *shared key reply* **then**
6:     ADD $u \in V$ to neighbor list as a shared key neighbor
7:     $i \leftarrow i + 1$
8: **else**
9:     ADD key ring $k_u$ and state of $u \in V$ to neighbor list
10:     $j \leftarrow j + 1$
11: **endif**
12: WAIT *neighbor discovery timer* out
13: **if** *straight_method* **then**
14:     **goto** END
15: **else if** *reactive_method* **then**
16:     **if** there is no *floating* neighbor in $j$ neighbors **then**
17:         **goto** END
18:     **else**
19:         BROADCAST key rings $k$ of only *floating* unsecured neighbors for *path key discovery*
20: **else if** *proactive_method* **then**
21:     BROADCAST key rings $k$ of all unsecured neighbors for path key discovery
22: **end if**
23: PATH KEY DISCOVERY:
24: RECEIVE path keys $K_{PK}$ from each $u_{1,2,...,I}$
25: **for** selected $j$ neighbors $u_j \in V$ of $v$ **do**
26:     **if** one $K_j$ exists in $K_{PK}$ **then**
27:         SEND $K_j$ to $u_j$ as the shared key
28:         ADD $u_j$ to neighbor list as a shared key neighbor
29:         $i \leftarrow i + 1$
30:         $j \leftarrow j - 1$
31:         *new_neighbor_added* $\leftarrow 1$
32:     **end if**
33: **end for**
34: **if** *proactive_method* **and** $j > 0$ **and** *new_neighbor_added* **then**
35:     BROADCAST key rings $k$ of all unsecured neighbors for path key discovery
36:     *new_neighbor_added* $\leftarrow 0$
37:     **goto** PATH KEY DISCOVERY
38: **end if**
39: END:

### 3.2 Elliptic Curve Diffie-Hellman (ECDH)

The cross layer implementation of ECDH key exchange over the configuration protocol [16] requires no additional messaging. The implementation is simple since the transaction between any node pair only requires the exchange of public keys $Pu$ to calculate the shared key for that link. As it is depicted in Fig. 3, *A* first broadcasts a polling message for neighbor discovery which includes also its public key $Pu_A$. Then, all the receiving neighbors $n_i$ ($i=1,2,...,d$) add their public keys $Pu_i$ in their reply messages. The bulk of the cost in this operation is the point multiplication of neighbor public key with the private key to calculate the link key for each neighbor sensor node.

We assume that the public keys are calculated offline by a powerful device and loaded to sensor nodes before the deployment. This operation saves one point multiplication operation needed for calculating the public key from private key in each sensor node. The analysis of the per node total energy cost of this key establishment scheme can also be divided into two parts as communication and computation costs.

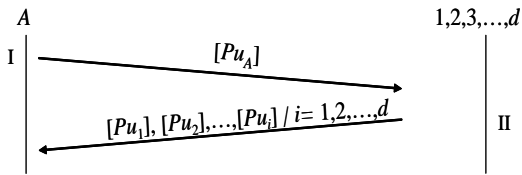

Fig. 3 Message transactions for ECDH.

### 3.2.1 Communication Cost of ECDH

The number of messages sent and received for ECDH key establishment is the same as the shared key discovery of BS since any node pair can establish a shared key with the probability of 1. Therefore, the message complexities defined in Lemma 1 and Lemma 2 are also valid for ECDH key establishment. Here, each node $v_i$ ($i = 1,2,3,\ldots,n$) transmits at most a single broadcast ($L_{poll}$) and $E_{neighbor}$ unicast ($L_{poll\_reply}$) messages. Furthermore, each node receives $E_{neighbor}$ broadcast ($L_{poll}$) and $E_{neighbor}$ unicast ($L_{poll\_reply}$) messages from its neighbors. The length of the messages used for security operations is 160 bits and it is the same for both broadcast and unicast. Consequently, the total messaging cost of the ECDH is calculated as below:

$$C_{M_{neighbor}} = C_{tx}(L_{poll} + E_{neighbor}L_{poll\_reply}) + \\ C_{rx}E_{neighbor}(L_{poll} + L_{poll\_reply}) \quad (15)$$

### 3.2.2 Computational Cost of ECDH

In ECDH key establishment protocol, total amount of computation is $E_{neighbor}$ point multiplications needed for establishing shared keys with the one hop distant neighbor nodes.

### 3.2.3 Total Energy Cost of ECDH

The message payload structure used in the energy cost analysis of the ECDH was:

$$A \rightarrow B : ID_A, ID_B, M, MAC\{ID_A, ID_B, M\}$$

The poll message ($L_{poll}$) contains the 160 bits public key of the source node. If the expected physical node degree $E_{neighbor}$ is $d$, then there are $d$ neighbors replying to this broadcast message. Each reply message ($L_{poll\_reply}$) includes the public key of the sender which is also 160 bits in length. After adding the 16 bits IDs and 160 bits MAC values, total messaging cost given in Eq. (15) can be rewritten as below:

$$C_{M_{neighbor}} = C_{tx}(160+192)(1+d) + C_{rx}(160+192)(2d) \\ = C_{tx}352(1+d) + C_{rx}704d \quad (16)$$

When the calculation cost per node is considered, the cost equation is simple since the public and private key pairs are generated by a powerful device and loaded before the deployment and there is only a single point multiplication computed for each node during the key establishment. If the number of physical neighbors is $d$, then the total computation cost is:

$$C_{computation} = d\, C_{ECDH_{160}} \quad (17)$$

The message length used in the SHA-1 MAC computation can be calculated by excluding the 160 bit MAC extension included in the communication cost defined in Eq. (16). So the total energy cost for SHA-1 computation is calculated as below:

$$C_{M_{SHA-1}} = C_{SHA-1}((160+32)(1+d) + (160+32)(2d)) \\ = C_{SHA-1}(192 + 576d) \quad (18)$$

Then, by using the equations Eq. (16), Eq. (17), Eq. (18) the overall per node energy cost of ECDH for $v_i$ ($i = 1,2,3,\ldots,n$) is calculated as below:

$$C_{configuration} = C_{M_{neighbor}} + C_{computation} + C_{M_{SHA-1}} \quad (19)$$

### 3.2.4 Cross Layer Implementation of ECDH

Since the links to be secured are controlled by the network configuration protocol there may be no need for *A* to establish a secure link with all of its neighbors for an acceptable network connectivity performance. Here, we can apply *reactive* and *proactive* link selection methods for ECDH also.

--For the *proactive* selection all the possible links among the neighbors are tried to be secured which will provide the same connectivity performance of the network configuration protocol without security implementation.

--With the *reactive* selection, only the links to the *floating* neighbors are tried to be secured.

Since the link key probability is 1 for the ECDH, if the physical node degree is selected as low as possible, then the computation cost could be reduced accordingly. This basic algorithm is provided in Table 5.

Table 5: ECDH cross layer algorithm

| |
|---|
| **Assumptions:** Network is $G = (V;E)$, there is no node failure during configuration |
| 1: BEGIN: |
| 2:     BROADCAST *poll* with $Pu_v$ for neighbor discovery and start *neighbor discovery timer* |
| 3:     /*proactive*: all neighbors reply; *reactive*: only *floating* neighbors reply*/ |
| 4:     RECEIVE *poll_reply* and $Pu_u$ from $u$ |
| 5:     $K_{uv} \leftarrow Pu_u \ P_u$ |
| 6: END: |

The simulation results of these implementations and their impacts on security and network configuration performance have been provided in Section 4.

### 3.2.5 Analysis of Signature Schemes for ECDH

The candidate protocols for implementing signature operations on ECDH key establishment protocol are RSA and ECDSA. Per unit energy costs of these protocols for Mica2dot sensor nodes are provided in Table 6 [11]. In this table, it can be realized that RSA signature generation operation is 13 times more costly than that of ECDSA protocol. However, the cost of RSA verification is about 3 times cheaper than ECDSA verification. Here, it can be assumed that the signature generation operations can be handled by a powerful device and the resulting signatures can be loaded to sensor nodes before deployment. In this case, RSA protocol is the most cost effective solution since its computational asymmetry is in favor of the signature verification.

Since ECDH is the most effective public key establishment protocol suitable for WSN [11], hybrid operation of ECDH and RSA can provide the most cost effective solution for authenticated key establishment using public key cryptography. Such a hybrid implementation has been proposed in [19] for mobile devices. On the other hand, using two different asymmetric protocols for one application adds another cost which is the program memory allocation. Since the signature verification operations are handled during the configuration, it is enough to implement just the public key operations of RSA in sensor nodes. In [18] it has been stated that the implementation of public key operations occupies only 1Kbyte of program memory for an 8 bits microcontroller. In that work, it has also been stated that the implementation of 160 bits ECC operations occupies 3,68 Kbyte of program memory for the same platform. Additionally, the maximum length of the RSA signature message equals to the RSA key length which is 1024 bits in our case [11]. Thus, the additional energy costs of the signature operations to the ECDH key establishment are the 1024 bit RSA signature message added to the transactions defined in Fig. 3 and the signature verification computation cost which is 50% of the ECDH key establishment.

Table 6: Asymmetric computation costs for Mica2dot

| Signature | Generation | Verification |
|---|---|---|
| RSA 1024 | 304mJ | 11,9mJ |
| ECDSA 160 | 22,82mJ | 45,09mJ |
| Key establishment | Client | Server |
| RSA 1024 | 15,04mJ | 304mJ |
| ECDSA 160 | 22,3mJ | 22,3mJ |

## 4. Simulation Results

In this section we provide the simulation results and performance comparisons of the cross layer implementations of key agreement protocols with the network configuration protocol. The simulation platform is based on Omnet++ [20] discrete event simulator. Results are averaged from 10 different runs each has a different random seed to create a different 2D uniform deployment. The maximum cluster size parameter of the network configuration protocol [16] has been set to 32.

The main purpose of the simulations is to measure the total number of bits to be transmitted and received and the number of computations to be made for key establishment. Using the measurement results, we calculated the total energy cost of transmission and computation based on the unit energy costs measured for Mica2dot sensors [10][11] provided in Table 2, 3 and 6. The resilience performance is calculated according to the following rules: a) When a node is compromised, all the secure links connected to this node and the other links that are using the keys stored in that node are assumed to be compromised, b) In BS, since path keys are encrypted with the shared keys stored in sensor nodes during transmission, if these shared keys used for encrypting the path keys are compromised, then the corresponding path keys are also compromised.

We have first investigated the performance of the cross layer implementation of the BS. The selected network configuration parameters for this simulation are: $n = 2000$, $d = 8$ and $P = 10.000$. For these parameters key index (*idx*) and node *ID* numbers are represented in 16 bits. We have measured the network connectivity, total number of messages and computations of straight, reactive and proactive link selection methods for different key ring sizes. Then, we calculated the overall energy costs based on the unit costs provided in Table 2 and 3.

In Fig. 4, we have provided the graphical representations of the analysis and simulation results of the total energy cost of BS for the given network configuration parameters. We have obtained the analysis results by multiplying the per node energy cost provided in Eq. (14) with the network size $n$. The simulation result has been obtained by implementing the BS to the network configuration protocol without applying the cross layer link selection algorithm. We have used *cascade-off counting* method in this simulation since the analysis was made for this case. From Fig. 4, it can be realized that the simulation and the analysis results display mainly the same characteristics for BS. In this figure, the analysis result of the energy cost is higher than that of simulation because analysis includes duplicated key establishment processes for some links that cannot be eliminated which are decided in real time during the configuration. This duplicated calculation increases the total energy cost of the analysis. However, in simulations, we can prevent nodes to establish duplicate keys for already configured links in real time. As a result, the simulation results have relatively lower energy costs. Hence, we can take the analysis result of the implementation of BS as an upper bound for the total energy cost calculated for *cascade-off counting*.

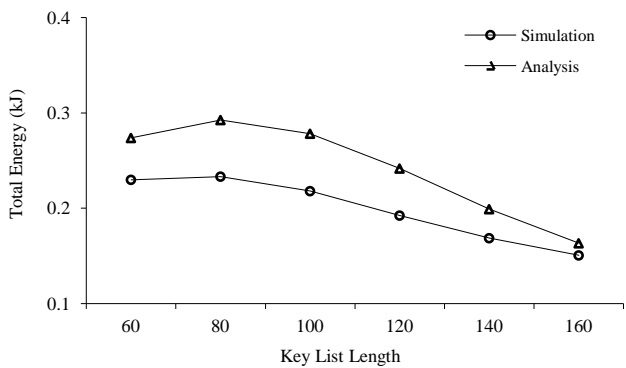

Fig. 4 BS cost analysis verification for $P = 10.000$.

From Fig. 5 to Fig. 7 we have provided the simulation results of the cross layer implementation of BS for the defined network parameters. In Fig. 5, the straight link selection requires a key ring of size 160 to achieve the 99% global network connectivity whereas, proactive and reactive methods need a much smaller key ring size (around 100) for the same level of global connectivity.

As it can be seen in Fig. 6, total energy cost of the key establishment is the lowest with the straight method. This result is due to the fact that straight method needs no messaging for the path key establishment. However, each sensor node should keep a key ring of size 160x128 = 20.480 bits (2.560 bytes) in its memory. Also, increasing the key ring size stored in each sensor node reduces the network resilience. On the other hand, eventhough the energy cost of the path key establishment is relatively higher, cost of the reactive method is about 13% less than that of proactive link selection for $k = 100$. Moreover, it provides almost equal global network connectivity (Fig. 5) and resilience (Fig. 7) compared with the proactive link selection. However, the overall resilience of the system is very low even for the smallest possible key ring size is selected for the configuration. For example, it is enough to capture 100 nodes from the network to compromise the 75% of the total links established with the proactive link selection method as shown in Fig. 7 (for $k = 100$). The reason of this low resilience is because of the selected key pool size. To increase the resilience, the key pool size $P$ should be increased to reduce the possibility of established link keys to be in the compromised set of keys.

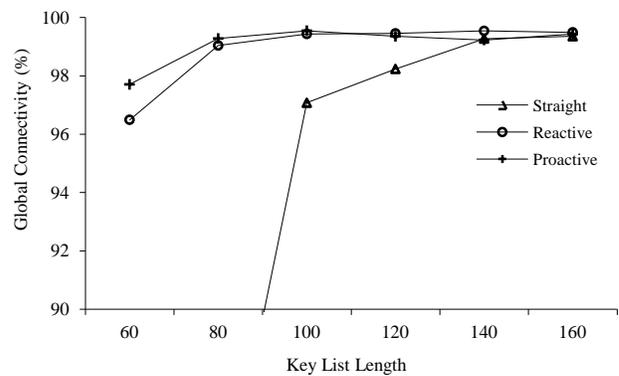

Fig. 5 BS global connectivity for $P = 10.000$.

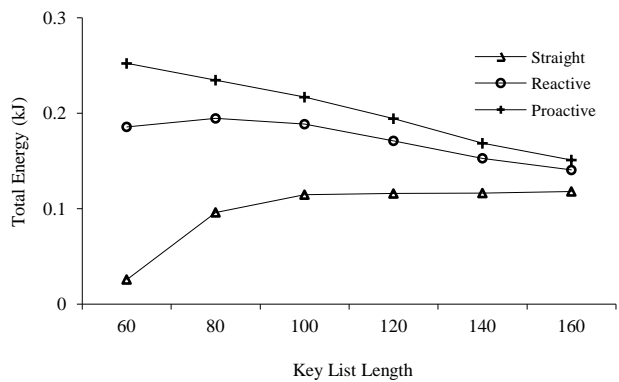

Fig. 6 BS configuration energy cost for $P = 10.000$.

We have also simulated the cross layer implementation of BS for enlarged key pool size $P = 100.000$ which is necessary to increase the overall resilience achieved with $P = 10.000$. However, increasing the key pool size will also cause the energy costs to increase because of the longer key rings to be transmitted and received for achieving the desired global connectivity. In Fig. 8, we have provided the global connectivity performance results for this simulation. Here, the key index can be represented in 17 bits. As it can be realized from the Fig. 8, the size of the key ring, required for the 99% global connectivity increases from 160 to 500 for the straight method. However, proactive and reactive methods can achieve this global connectivity with the key ring size around 300 instead of 100. Hence, increasing the key pool size increases the required key ring size to be stored in sensor memory. This also increases the total communication costs especially during the path key discovery. As shown in Fig. 9, total energy costs for $P = 100.000$ almost triples the costs for $P = 10.000$ provided in Fig. 6. The total energy cost of the reactive method is again around 15% less than that of proactive method because of the reduced path key establishment rounds together with the cross layer link selection.

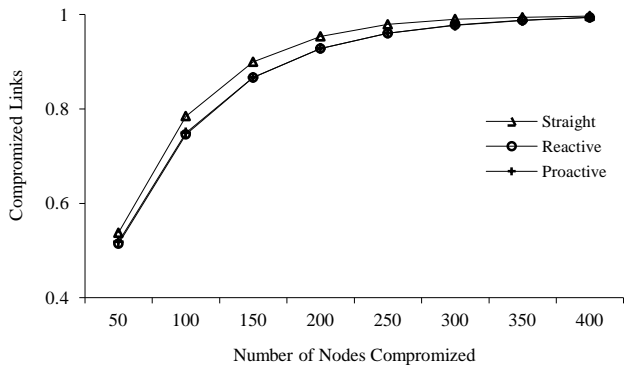

Fig. 7 BS resilience for $P = 10.000$.

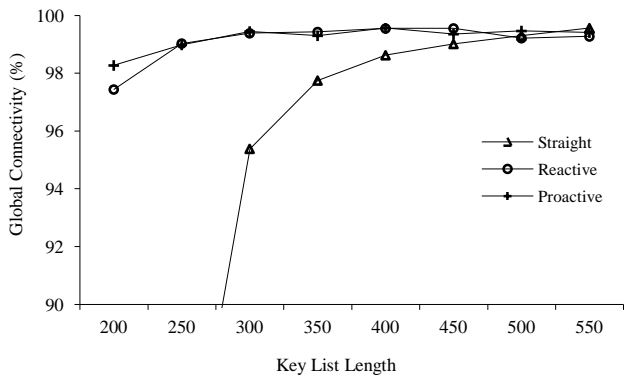

Fig. 8 BS global connectivity for $P = 100.000$.

On the other hand, the overall resilience of the system is increased considerably as shown in Fig. 10. In this case, it is needed to capture around 300 nodes from the network, which was 100 for $P = 10.000$, to compromise the 75% of the total links established with the proactive link selection ($k = 300$). This resilience characteristic is the same for both proactive and reactive methods which is better than the value 81% achieved with the straight method. The reason of this improvement is the smaller key rings stored in each node for proactive and reactive methods which reduce the probability of including established link keys in these key rings.

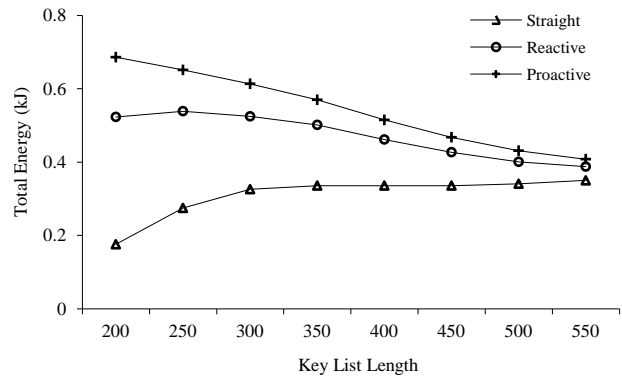

Fig. 9 BS configuration energy cost for $P = 100.000$.

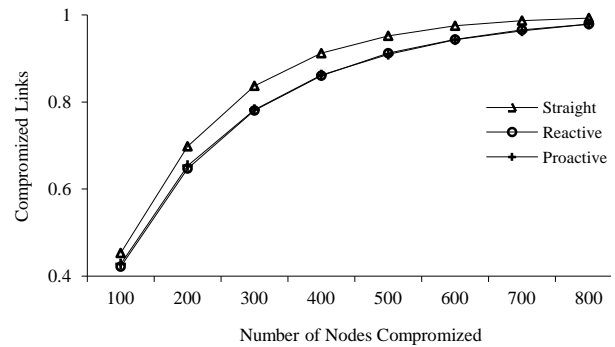

Fig. 10 BS resilience for $P = 100.000$.

We observed from the simulations that the overall energy cost of BS comes mostly from the radio transmissions and the SHA-1 MAC calculation costs. To achieve a better resilience the key pool size can be increased more, but in this case the number of keys to be stored in each of the sensor nodes has to be increased for the desired local connectivity. Then, the total energy cost increases because of the transmission of long key ring indexes.

The performance of the cross layer implementation of ECDH key establishment is provided in Table 7.

Table 7: ECDH key agreement costs

| Link Type | Connectivity | Energy |
|---|---|---|
| Proactive | 99,34% | 0,42 kJ |
| Reactive | 99,25% | 0,40 kJ |

For this simulation the selected network configuration parameters are: $N = 2000$ and $d = 8$. Here, node *ID* numbers are represented in 16 bits. We have measured the network connectivity and total number of messages and computations of reactive and proactive link selection methods. Then, we calculated the overall energy costs based on the unit costs provided in Table 2, 3 and 5.

Simulation results indicate that the connectivity performances of the proactive and reactive methods are also the same for ECDH which are above 99%. In this case, the total energy cost of the reactive method is about 5% less than that of the proactive method. The ECDH energy cost simulation results given in Table 7 matches with the 0,43 kJ computed using Eq. (19). The small difference is because of the duplicated message costs counted in the analysis. The interesting result here is that the total energy cost of the ECDH implementation to achieve 99% network connectivity is about 20% less than that of BS ($P = 100.000$ and $k = 300$) for both proactive and reactive link selection methods. As shown in Fig. 11, the resilience of the ECDH is also much better than BS such that to compromise the 75% of the total secure links, 1000 nodes are needed to be compromised from the network.

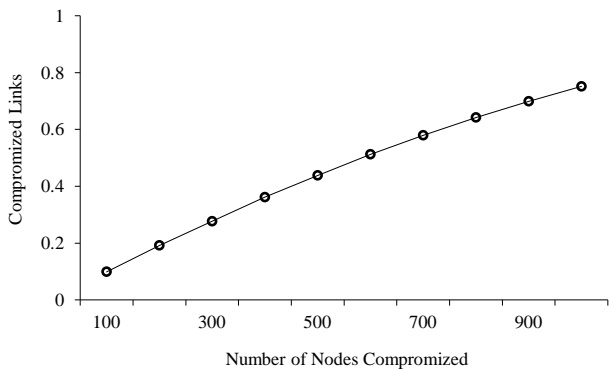

Fig. 11 Resilience performance of ECDH for reactive method.

When we consider the network connectivity and total energy cost characteristics with the increasing network size, Fig. 12 provides the simulation results of the BS and ECDH running in reactive mode. Here, we have selected key pool size as $P = 100.000$ and key ring size as $k = 300$ for BS. The selected node degree of the network is the same with the previous test beds. In Fig. 12-(a) the network connectivity performance is almost the same for both implementations with only small differences. The global connectivity is below 99% with the network size of 400 because of the border conditions but it is always above 99% for the sizes starting from 1200. Thus, it can be stated here that the connectivity results of the key establishment protocol implementations are independent from the network size.

The energy cost characteristics provided in Fig. 12-(b) illustrate that as the network size increases, energy costs increase linearly for both key establishment protocols.

However, the rate of increase for the BS is about 30% higher than that of ECDH. Consequently, reactive cost of ECDH is over performing the reactive cost of BS in terms of the overall network energy usage when establishing keys among sensor nodes while keeping the network configuration performance at an acceptable level.

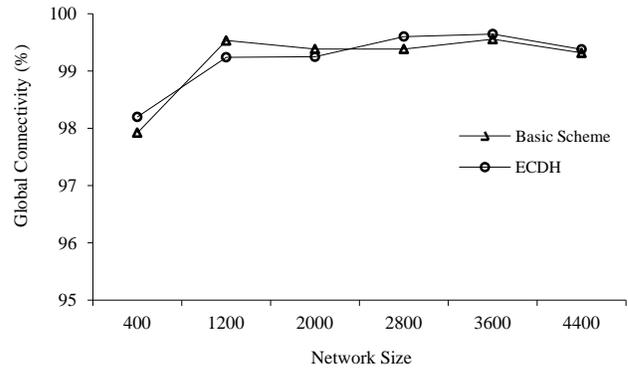

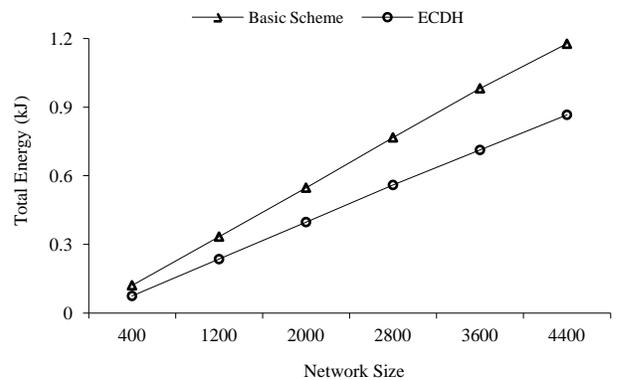

Fig. 12 (a) Network connectivity and (b) Energy cost characteristics of reactive BS and ECDH for the increasing network size.

The simulation results so far reflect the analysis of different BS and ECDH key establishment setups over the network configuration protocol [16] when the physical node degree is 8. For this node degree the key ring size should be enlarged for BS so that the local connectivity is close to the physical node degree to achieve the global connectivity of 99%. On the other hand, for ECDH, there is no special configuration change required to fulfill the global and local connectivity requirements. We have observed from these results that with the key pool size of $P = 10.000$ BS always has the lower energy costs compared to ECDH but the achieved resilience is just 10% of ECDH. If the key pool size is increased to 100.000, resilience performance of BS is still below ECDH but it increases considerably. However, BS loses its energy cost advantage over reactive link selection with ECDH especially with the reactive link selection (for example if $k = 300$). At this point, one can propose to use the straight method with higher key ring sizes ($k \geq 500$) which keeps the energy cost advantage for BS but in this case resilience levels become worse because of the increased key ring size as shown in Fig. 10.

In addition, ECDH resilience depends only on the network size but BS resilience depends on key pool $P$ and key ring $k$ sizes. That is if network size is increased while keeping $P$ and $k$ the same, BS resilience characteristic will be the same as in Fig. 10 but ECDH resilience changes. For example, if the network size is increased to $n = 20.000$, compromising 1000 nodes results in compromising whole network for BS but for ECDH only 10% of the links are compromised. This is the main advantage of ECDH over BS especially for the large scale networks.

In order to complete the cost characteristics of BS and ECDH implementations over the network configuration protocol we have investigated the energy cost characteristics when the physical node degree increases. Table 8-a provides the simulation results of the energy costs of BS for the physical node degrees of $d = \{8, 13, 18\}$ for the key pool size $P = 100.000$ and the global connectivity above 98%. For this simulation we have set the network size as $n = 400$ because of the simulator limitations and due to the border conditions we have set the global connectivity limit to 98%. For the comparison we have used the reactive ECDH energy cost 0,074kJ which is calculated for $d = 8$ and $n = 400$. Here, this energy cost is assumed to be fixed for higher node degrees as the total number of neighbors securely connected with ECDH is 8 and it does not change. Increasing the physical node degree adds extra polling messages to the overall cost but this messaging cost is negligible when it is compared to the total ECDH energy costs. From the results in Table 8-a, straight link selection cost of BS almost equals to the ECDH key establishment cost for $d = 8$ but reactive link selection energy cost of BS is higher. For $d = 13$, BS straight link selection cost is 27% less than ECDH reactive cost and it becomes feasible against ECDH at this point. The decrease in BS straight link selection energy cost continues with $d = 18$ but the decrease rate is slower due to the closer key ring lengths achieving the same connectivity performances for $d = 13$ and $d = 18$. However, BS reactive link selection cost increases as the node degree increases. The reason is that, as the physical node degree increases, the number of neighbors participating to the phase key discovery phase of BS increases (all physical neighbors) and this increases the total communication cost. However, in the straight link selection only the neighbors having shared key in their key rings responds and this removes the unnecessary communication with other neighbors. Actually, this advantage is not valid for $d = 8$ because each node needs to connect to all of its neighbors for the desired network connectivity performance. However, for $d > 8$ only a portion of the physical neighbors is enough and there is no need to establish a secure link with the others which results in a reduction in the total energy cost.

We have repeated the simulations for $d = 13$ and $d = 18$ by setting the key ring size as $k = 500$ for the BS straight link selection and we have obtained the total energy costs 0,08kJ and 0,11kJ respectively. Then ECDH simulations have been repeated by allowing secure link connection with all the physical neighbors without any limitation and then we have obtained the total energy costs of 0,096kJ and 0,14kJ respectively. As it can be realized from Table 8-b, if the network configuration requires maximum local connectivity achievable when $d = 18$, BS straight selection cost is becoming about 20% lower than the ECDH reactive cost. This difference is very low when $d = 8$ which is ignorable. On the other hand, for the BS reactive selection, energy costs are increasing considerably to the levels of 0,2kJ and 0,3kJ. From these results it is clear that for the node degrees $d \geq 13$, Straight link selection with BS has the advantage over reactive implementation of ECDH. However, for the node degree of $d = 8$, straight link selection cost of BS almost equals to the ECDH reactive cost. Thus, because of its better resilience ECDH would be preferred for the node degree level of $d = 8$. But for the higher node degrees straight link selection with BS would be the choice because of its lower energy cost.

Table 8: Energy costs versus increasing physical node degree

*(a) Minimum Local Connectivity*

| BS $d = 8$ | Straight | Reactive |
|---|---|---|
| Key ring size | 500 | 300 |
| Energy cost | 0,071kJ | 0,092kJ |
| BS $d = 13$ | Straight | Reactive |
| Key ring size | 300 | 160 |
| Energy cost | 0,055kJ | 0,155kJ |
| BS $d = 18$ | Straight | Reactive |
| Key ring size | 250 | 100 |
| Energy cost | 0,048kJ | 0,245kJ |
| ECDH | Straight | Reactive |
| Energy cost | --- | 0,074kJ |

*(b) Maximum Local Connectivity*

| BS $d = 8$ | BS Straight | BS Reactive | ECDH Reactive |
|---|---|---|---|
| Key ring size | 500 | 300 | --- |
| Energy cost | 0,071kJ | 0,092kJ | 0,074kJ |
| BS $d = 13$ | BS Straight | BS Reactive | ECDH Reactive |
| Key ring size | 500 | 300 | --- |
| Energy cost | 0,08kJ | 0,21kJ | 0,096kJ |
| BS $d = 18$ | BS Straight | BS Reactive | ECDH Reactive |
| Key ring size | 500 | 300 | --- |
| Energy cost | 0,11kJ | 0,33kJ | 0,14kJ |

Finally, Table 9 provides the energy cost results of the ECDH & RSA authenticated key establishment operations for $n = 2000$ and $d = 8$. According to these results, the reactive link selection cost of the ECDH & RSA hybrid implementation is 50% higher than that of ECDH only key establishment. However, this cost is only 20% higher than that of BS for the given configuration parameters (reactive selection with $P = 100.000$ and $k = 300$) and it is paid only once in the beginning of the configuration. Thus, this additional cost can be affordable for the sensor networks requiring strong security services since it provides a well proven authentication and a very good resilience compared to BS.

Table 9: ECDH and RSA signature hybrid implementation costs

| Link Selection | $ECDH_{160} + RSA_{1024}$ |
|---|---|
| Proactive | 0,71kJ |
| Reactive | 0,66kJ |

## 5. Conclusions

In this work we have provided the design and analysis of the implementations of BS [2] and ECDH [9] over the network configuration protocol proposed in [16] and measured the performances under different configuration parameters. By using the cross layer relations between key establishment and network configuration protocols, we have proposed three different link selection methods, straight, reactive and proactive, for securing the links at the configuration time. We have simulated the performance of these cross layer implementations for different setups by changing network size, key pool size, key ring size and node degrees.

Simulation results indicate that the cost of the key establishment protocols can be improved by using the state control mechanism provided by the network configuration protocol. The cost of BS can be reduced about 15% by selectively securing the links at configuration time. This reduction is 5% for ECDH. When the key pool size is 10.000 then the total energy cost of the BS is always better than ECDH but the resilience performance is only 10% of ECDH. For the minimum local connectivity target ($d = 8$), which provides 99% global connectivity for the network configuration protocol, if the key pool size is increased to 100.000 then the resilience performance of BS becomes closer to the performance of ECDH but its total energy cost increases due to the increase in the total transmission cost. For instance, when the key ring of size is 500, straight link selection cost of BS is almost equal to the ECDH reactive cost and BS reactive link selection cost is 25% higher for the key ring size of 300.

Simulation results investigating the node degree and energy cost relations of BS and ECDH over the network configuration protocol indicate that reactive link selection cost of ECDH is in range of straight link selection cost of BS if the node degree is set to the minimum ($d = 8$). If the node degree increases, then BS and ECDH energy costs also increase but straight link selection cost of BS becomes 20% lower than the ECDH reactive cost and the reactive link selection cost of BS almost doubles. Then, it can be stated here that for the minimum node degree requirement of the network configuration protocol ($d = 8$), ECDH is a good choice for the key establishment. For higher node degrees BS has a better energy cost performance and can be the preferred choice.

On the other hand, the resilience that can be achieved by BS is not as good as ECDH even with the key pool size of 100.000. To increase the resilience, key pool size can be increased more but this causes the key rings to be larger which results in higher energy costs. Besides, because of the key rings needed to be stored in each sensor node, BS suffers from the local storage overhead. The path key discovery phase can reduce this overhead but it also increases the transmission cost. So if the security is the prime concern and requires high resilience performance, then ECDH still can be a good choice even for the higher node degrees.

In addition, it should be noted that increasing the node degree will increase the total system cost especially for large scale WSN. Since the network configuration protocol allows 99% global network connectivity with the node degree of $d = 8$ this would be the best configuration parameter to build the network. Then ECDH is becoming the best choice for this configuration setup. On the other hand, cross implementation of ECDH is fairly easy since each communicating party can establish a link key directly. For this case, each node needs to store only public and private keys having total size of only 40 bytes which is ignorable. In addition, hybrid implementation of ECDH and RSA can be used for authenticated key establishment in return for an acceptable cost and this operation could provide even more security by adding the pairwise authentication to the key establishment.

As a result, the cross layer implementation of key establishment with the network configuration can give more realistic performance results and there can be improvements on operational costs by reducing the number of links to be secured without affecting the configuration targets. For the optimum configuration parameters of the network configuration protocol [16] ECDH is the best performing key establishment protocol compared to BS. Moreover, ECDH and RSA hybrid implementation can be used for higher security needs. These results can be generalized to the network configuration protocols which can achieve a global connectivity of 99% with the physical node degrees of around 8.

**Özgür SAĞLAM** received his B.S. degree from the Electrical and Electronics Engineering Department, 9 Eylül University, İzmir, Turkey, in 1998. He then received the M.S. and Ph.D. in International Computer Institute from Ege University, İzmir, Turkey, in 2002 and 2009 respectively. Currently he is a project manager of R&D Department at Arcelik A.Ş. Electronics Plant. His research interests include distributed networks, network security and wireless sensor networks.

**Mehmet E. DALKILIÇ** received his B.S. degree from the Electrical and Electronics Engineering Department, Hacettepe University, Ankara, Turkey, in 1985. He then received the M.S. and Ph.D. in Computer Engineering from Syracuse University, New York, USA, in 1989 and 1994 respectively. Currently he is a professor of Computer Science at Ege University and director of the International Computer Institute. His research interests include algorithm design, networks and security.